\documentclass[10pt,aps,prb,twocolumn,floatfix,amssymb]{revtex4-1}
\usepackage{graphicx,subfigure}
\usepackage{bm}
\usepackage{amsfonts}
\usepackage{dsfont}
\usepackage{color}

\begin{document}
\title{Phase diagram of the isotropic spin-3/2 model on the $z=3$ Bethe lattice}
\author{Stefan Depenbrock}
\email{stefan.depenbrock@lmu.de}
\affiliation{Department of Physics and Arnold Sommerfeld Center for Theoretical Physics, Ludwig-Maximilians-Universit\"at M\"unchen, 80333 M\"unchen, Germany}

\newcommand{\frank}[1]{ { \color{red} \footnotesize (\textsf{FP}) \textsf{\textsl{#1}} } }
\newcommand{\stefan}[1]{ { \color{blue} \footnotesize (\textsf{SD}) \textsf{\textsl{#1}} } }

\author{Frank Pollmann}
\email{frankp@pks.mpg.de}
\affiliation{Max-Planck-Institut f\"ur Physik komplexer Systeme, 01187 Dresden, Germany}

\begin{abstract}
We study an  $SU(2)$ symmetric spin-3/2 model on the $z=3$ Bethe lattice using the infinite Time Evolving Block Decimation  (iTEBD) method. This model is shown to exhibit a rich phase diagram. We compute the expectation values of several order parameters which allow us to identify a ferromagnetic, a ferrimagnetic, a anti-ferromagnetic as well as a dimerized phase. We calculate the entanglement spectra from which we conclude the existence of a symmetry protected topological phase that is characterized by $S=1/2$ edge spins. Details of the  iTEBD algorithm used for the simulations are included.
\end{abstract}
\maketitle

\section{Introduction}

Quantum antiferromagnets are a challenging and interesting topic in condensed matter physics. The interplay between strong correlations, quantum fluctuations, and frustration can yield very exciting new phases with often unexpected  properties.  One dimensional quantum spin chains have been proven to be very useful to understand many of these interesting phases. For these systems, very powerful analytical and numerical methods exist. A major advance on the numerical side was the introduction of the Density-Matrix-Renormalization Group (DMRG) method which allows for the efficient simulation of one-dimensional systems.\cite{White-1992}  One of the first successes of the then new DMRG method was its use to prove the famous Haldane conjecture for integer spin chains numerically. The Haldane conjecture states that the Heisenberg antiferromagnetic (HAF) chain with  integer spin $S$ has a nonzero excitation gap and exponentially decaying spin correlation functions (while spin chains with half-integer spin are gapless).\cite{Haldane-1983a,Haldane-1983} For odd integer spins, the Haldane phase  is an example of a so-called \emph{symmetry protected topological phase} (SPTP) which is characterized by $S=1/2$ edge spins. This kind of phase cannot be characterized by symmetry breaking but instead by using cohomology theory.\cite{GuWen-TEFR2009, Chen-2011,Chen-2011a,Pollmann-2012,Pollmann-2010,Schuch-2011}

In this paper we make use of an extension of the DMRG algorithm to the Bethe lattice and study the phase diagram of a general $SU(2)$ symmetric $S=3/2$ spin model on a Bethe lattice with coordination number $z=3$.
This  model has a special point in this phase diagram, the so-called Affleck-Kennedy-Lieb-Tasaki (AKLT) point, at which the ground state is known exactly.\cite{Affleck-1987} The AKLT wave function for this model  is a quantum paramagnet with exponentially decaying correlation functions.\cite{Affleck-1987, Fannes-1992,Niggemann-1999,Laumann-2010}  
We argue that the AKLT point extends to a phase which is similar to the Haldane phase in one-dimensional spin chains in that it is characterized by a fractionalized edge spin (the precise meaning of an edge spin in this context is explained Sec.~\ref{sec:sptp}).  
Beside the Haldane phase, the model is shown to exhibit different magnetic phases as well as a dimerized phase. 
Even though there exist experimental systems, such as dendrimers that realize the tree structure\cite{Tomalia-1994} related to the Bethe lattice considered in this work, we are mainly interested in this system because of its theoretical nature.
The $S=3/2$ model has already  been shown to exhibit a very complex phase diagram on the mean field level and in one-dimensional systems. \cite{Wu-2006,Fridman-2011,Rachel-2013}  
The main goal is to present a conclusive phase diagram of the model utilizing recently introduced algorithms which allow an efficient simulation of quantum spin systems on Bethe lattices. 
We use a descendant of Vidal's infinite time-evolving block decimation algorithm (iTEBD) \cite{Vidal-2007} adapted to the tree like structure. The  iTEBD method as well as the DMRG have already successfully been applied to reproduce, e.g., the phase diagram of the transverse field Ising model and the spin 1/2 XXZ model on the $z=3$ Bethe lattice.\cite{Delgado-2002,Nagaj-2008,Nagy-2012,Manoranjan-2012,Li-2012} We include  details of the algorithm in this paper and discuss a number of improvement that make it more stable.


%
%

This paper is organized as follows: in Sec.~\ref{sec:mod} we introduce the model and discuss some of its basic
properties leading up to section Sec.~\ref{sec:method} where we describe the method we used to obtain the results. In Sec.~\ref{sec:sptp} we first take a closer look at symmetry protected topological phases before we present the results in Sec.~\ref{sec:results}. The key points of this paper are summarized again in Sec.~\ref{sec:sum}.

\section{Model\label{sec:mod}}
Throughout this paper we consider the following nearest-neighbor spin-3/2 model Hamiltonian
\begin{equation} \label{eq:model}
H=  \sum_{i} \alpha \vec{S}_i \cdot \vec{S}_{i+1} + \beta \left(\vec{S}_i \cdot \vec{S}_{i+1} \right)^{2} + \gamma \left(
\vec{S}_i \cdot \vec{S}_{i+1} \right)^3 \label{H0}
\end{equation}
where a different parametrization of the Hamiltonian is given by
\begin{eqnarray}
\alpha &=& \cos \varphi \cos \theta \\
\beta &=& \sin \varphi \cos \theta \\
\gamma & =& \sin \theta
\end{eqnarray}
with $\varphi \in \left[- \pi, \pi\right]$ and $\theta \in \left[-\frac{\pi}{2}, \frac{\pi}{2}\right]$.
The symmetries of this model include translation, spatial inversion, $SU(2)$,  and time reversal (TR). This model is known to exhibit an AKLT-like
wavefunction at the point $\alpha = 1.0$, $\beta = \frac{116}{243}$, $\gamma = \frac{16}{243}$ when placed on a $z=3$ Bethe lattice. The AKLT state has symmetry protected $S=1/2$ edge spins which are discussed in detail in Sec.~\ref{sec:sptp}. Furthermore this model exhibits a $SU(4)$ symmetry at four points in the phase diagram connected by a $SO(5)$ symmetric line which is given by\cite{Wu-2006,Fridman-2011} $\alpha = - \frac{1}{96}(31J_0 + 23 J_2)$, $\beta = \frac{1}{72}(5J_0 + 17J_2)$, and $\gamma = \frac{1}{18}(J_0 + J_2)$ with $J_0, J_2 > 0$.
We place this Hamiltonian on the Bethe lattice with coordination number $z=3$. 
The Bethe lattice and its finite counterpart, the Cayley tree, has first been used in statistical mechanics.\cite{Falk-1975,Chakraborty-1984,Thouless-1986} 
More recently it has also proved to be a highly instructive testing ground for tensor network methods as it is loop-free thus removing one of the major sources of entanglement. 
Additionally, the Bethe lattice is self-similar, enabling the application of efficient infinite-system methods. 
This lattice is infinite by definition and thus there are no surface effects.\cite{Ostilli-2011} 
Note that the thermodynamic limit of the Cayley tree and the Bethe lattice are not equivalent.    
This inequivalence is rooted in the large number of surface sites contained in any Cayley tree.\cite{Changlangi-2012} 
Whereas in most systems the ratio of boundary to bulk sites reduces to zero for large systems, it remains finite for the Cayley tree. 
The finite tree is thus dominated by the boundary conditions, making it unsuitable to study the model's properties in the thermodynamic limit. 

\section{Tensor Product State Based Simulations on a Tree Lattice} \label{sec:method}
\subsection{Definitions}
For our simulations we  use the infinite tree tensor network state (iTTN) \cite{Li-2012,Shi-2006,Nagaj-2008,Tagliacozzo-2009,Murg-2010} representation  of the ground state wave function. The iTTN states are the natural choice of ansatz state for our model system, since they model the tree's geometry. We thus employ this representation to compute the ground state properties numerically, using the infinite time-evolving block decimation (iTEBD) method \cite{Vidal-2003,Vidal-2007,Orus-2008} adapted to infinite trees. The iTEBD method is a descendant of the density matrix renormalization group (DMRG) method \cite{White-1992,Schollwock-2005, Schollwock-2011} based on matrix product states (MPS) \cite{Fannes-1992} which can be generalized to trees. For the sake of completeness, we now review some of the properties of MPS's, followed by an introduction to TTN states. 

A translationally invariant MPS for a chain of length $L$ can formally be written in the following form
\begin{equation}
\left\vert \psi \right\rangle =\sum_{\{m_{j}\}}\text{tr}\left[\Gamma _{m_{1}}\Lambda \dots \Gamma _{m_{L}}\Lambda
\right]|m_{1}\dots m_{L}\rangle \text{.}  \label{mps}
\end{equation}
Here, $\Gamma _{m}$ are $\chi \times \chi$ matrices with $\chi$ being the dimension of the matrices used in the
MPS. The index $m=-S,\dots ,S$ is the \textquotedblleft physical\textquotedblright\ index, e.g., enumerating the spin
states on each site, and $\Lambda $ is a $\chi \times \chi $, real, diagonal matrix. Ground states of one dimensional
gapped systems can be efficiently approximated by matrix-product states\cite{Hastings-2007, Gottesman-2009,
Schuch-2008}, in the sense that the value of $\chi $ needed to approximate the ground state wavefunction to a given
accuracy converges to a finite value as $N\rightarrow \infty$. We therefore think of $\chi$ as being a finite (but
arbitrarily large) number, which can be used to control the simulation's precision.

The matrices $\Gamma $, $\Lambda $ can be chosen such that they satisfy the canonical conditions for an infinite
MPS\cite{Vidal-2003a, Orus-2008}
\begin{equation}
\sum_{m}\Gamma^{\vphantom{\dagger }}_{m}\Lambda ^{2}\Gamma _{m}^{\dagger
}=\sum_{m}\Gamma _{m}^{\dagger }\Lambda ^{2}\Gamma^{\vphantom{\dagger }}
_{m}=\mathds{1}\text{.}  \label{canonical}
\end{equation}
These equations can be interpreted as stating that the transfer matrix
\begin{equation}
T_{\alpha \alpha ^{\prime };\beta \beta ^{\prime }}=\sum_{m}\Gamma _{m\beta}^{\alpha }\left(\Gamma _{m\beta ^{\prime
}}^{\alpha ^{\prime}}\right)^{\ast}\Lambda _{\beta }\Lambda _{\beta ^{\prime }}\label{transfer}
\end{equation}
has a right eigenvector $\delta _{\beta \beta ^{\prime }}$ with eigenvalue $\lambda = 1$. ($^\ast$~denotes complex
conjugation.) Similarly, $\tilde{T}_{\alpha \alpha ^{\prime };\beta \beta ^{\prime }}=\sum_{m}(\Gamma_{m\beta ^{\prime
}}^{\alpha ^{\prime }})^{\ast }\Gamma _{m\beta }^{\alpha}\Lambda _{\alpha }\Lambda_{\alpha ^{\prime }}$ has a left
eigenvector $\delta _{\alpha \alpha ^{\prime }}$ with $\lambda =1$. We further require that $\delta _{\alpha \alpha
^{\prime }}$ is the \emph{only} eigenvector with eigenvalue $\left\vert \lambda \right\vert \geq 1$ (which is equivalent
to the requirement that $|\psi\rangle$ is a pure state\cite{PerezGarcia-2008}).

The considerations given here become most intuitive when one considers,
formally, an infinite chain. 
We form a partition of the chain by cutting a bond which results in two half-chains. The wavefunction can then be Schmidt decomposed \cite{Schmidt-1907} in the form
\begin{equation}
|\psi \rangle =\sum_{\alpha }\lambda _{\alpha }|\alpha L\rangle |\alpha
R\rangle ,  \label{schmidt}
\end{equation}
where $|\alpha L\rangle $ and $|\alpha R\rangle $ ($\alpha =1,\dots ,\chi $) are orthonormal basis vectors of the left
and right partition, respectively. In the limit of an infinite chain, and under the canonical conditions (\ref
{canonical}), the Schmidt values $\lambda _{\alpha }$ are simply the entries of the $\Lambda $ matrix, $\Lambda
_{\alpha \alpha}$. The  $\lambda _{\alpha }^{2}$ are the eigenvalues of the reduced density matrix of either of the two
partitions, and are referred to as the \emph{entanglement spectrum}. The entanglement entropy is $S=-\sum_{\alpha
}\lambda _{\alpha }^{2}\ln \lambda _{\alpha }^{2}$. This is the von Neumann entropy of the reduced density
matrix. The states $|\alpha L\rangle$ and $|\alpha R\rangle$ can be obtained by multiplying together all the matrices to
the left and right of the bond, e.g., if the broken bond is between sites $0$ and $1$,
$|\alpha L\rangle=\sum_{\{m_{j}\},j\leq0}\left[\prod_{k\leq 0} \Lambda \Gamma_{m_{k}} \right]_{\gamma\alpha} |\dots
m_{-2}m_{-1}m_0\rangle$. Here, $\gamma $ is the index of the row of the matrix; when the chain is infinitely long, the
value of $\gamma$ affects only an overall factor in the wavefunction. Reviews of MPS's as well as the canonical form can
be found in Refs.~\onlinecite{PerezGarcia-2007,Orus-2008}.

\begin{figure}
   \subfigure[\label{fig:2siteAnsatz}]{
   \includegraphics[width=110pt]{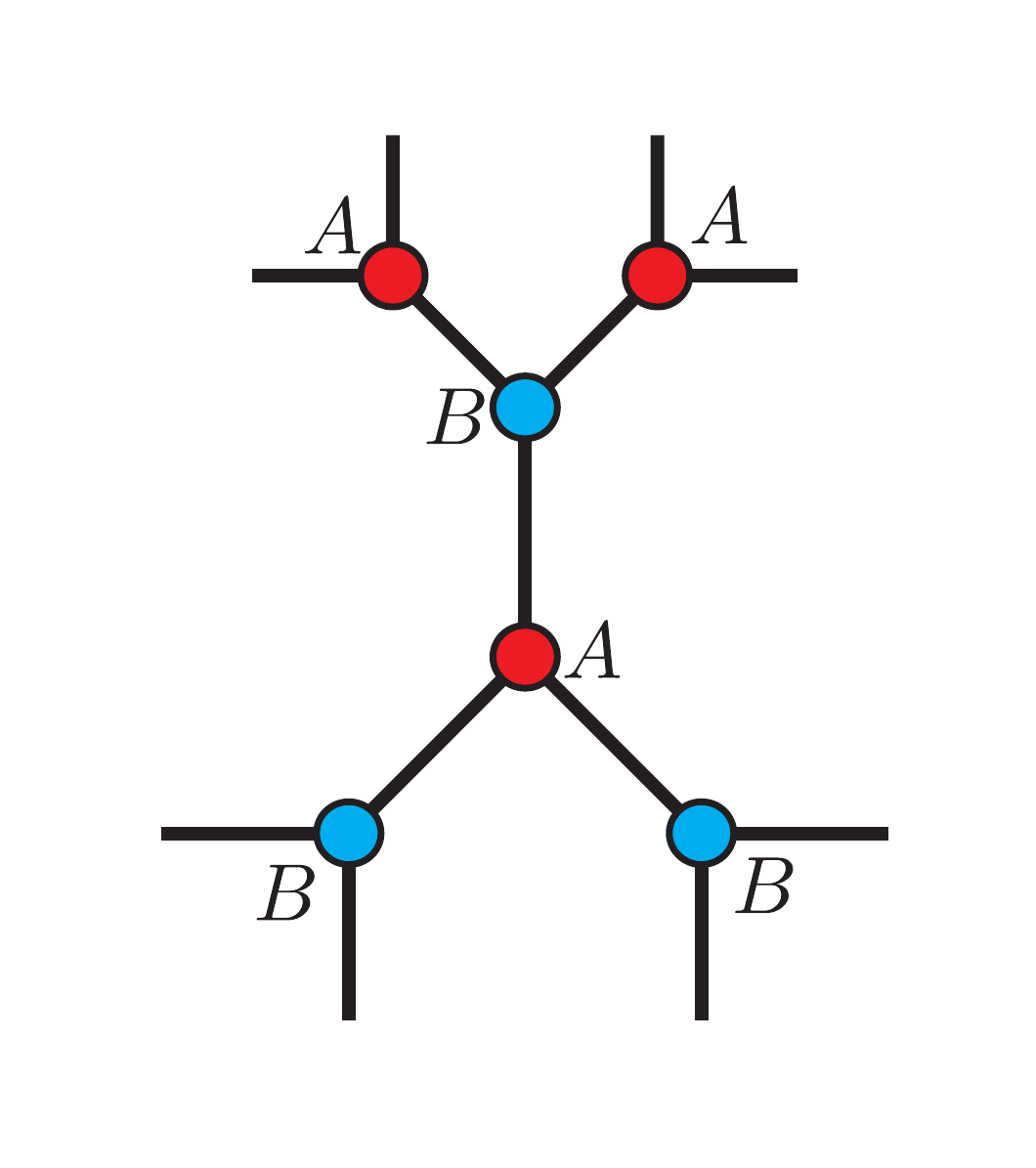}
   }
   \subfigure[\label{fig:6siteAnsatz}]{
   \includegraphics[width=110pt]{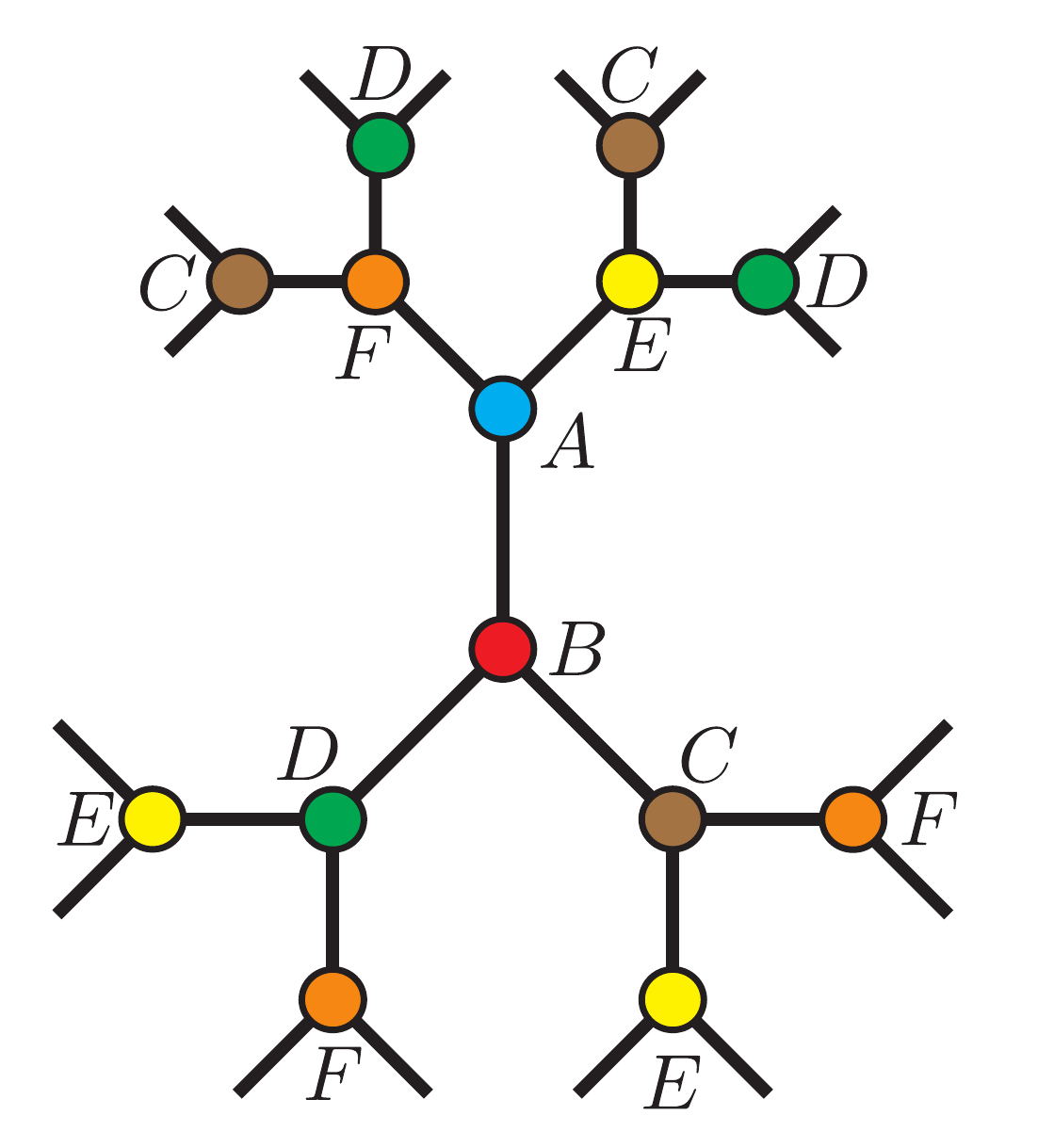}
   }
\caption{\label{fig:Ansatz}The two different tilings for the infinite Bethe lattice that were employed in the simulations.}
\end{figure}

\subsection{Tree Tensors Networks}
While MPS are the natural choice of ansatz state for one-dimensional systems, they are eminently unsuitable for large
higher-dimensional systems. For these systems the ideas behind MPS can be generalized to create a new class of states known as
tensor product (TPS) or projected entangled pair-states (PEPS).\cite{Verstraete-2004,Vidal-2008} The construction of these states is based on the bipartite nature of trees, allowing them to be split into two subsystems via the Schmidt decomposition, analogous to one-dimensional chains. Thus the generalization of the one-dimensional construction to trees is straightforward but in order to introduce our ansatz and the notation, we will cover it here as well.

To describe a tree of coordination number $z$ (i.e. each vertex has $z$ nearest neighbors), we place tensors $\Gamma^{[i]}$ of order $z+1$ on the vertices and vectors $\Lambda^{k}$ on the edges of the tree graph in Fig. \ref{fig:Ansatz}. We then connect the tensor's indices in a way that mimics the model's underlying lattice structure. A state $\vert \psi \rangle$ on the $z=3$ Bethe lattice can in this representation be written as
\begin{eqnarray*}
\vert \psi \rangle = &\left( \prod_{k \in \text{bonds}} \sum_{a_k < \chi} \Lambda^{k}_{a_k} \right)& \times \\
 &\left(\prod_{i \in \text{sites}} \sum_{s_i < d} \Gamma_{a_l a_m a_n}^{[i] s_i} \right)& \vert \dots \rangle \vert s_i
\rangle \vert \dots \rangle.
\end{eqnarray*}
While the dimension $d$ of the physical indices $s_i$ is dictated by the model, the dimension $\chi$ of the virtual indices $a_k$ can be chosen arbitrarily and is only limited by computational resources. This ansatz can easily be extended to lattices with a higher coordination number but in this publication we will only cover the case of $z=3$.  Analogous to MPS, the tensors in a tree tensor network can be chosen such that they satisfy the conditions for a canonical tensor network:
\begin{eqnarray}
&\sum_{a_k}& \Lambda^{2}_{a_k} = 1 \\
&\sum_{s_i}& \sum_{a_k a_l} \Gamma_{a_k a_l a_m}^{s_i} \Lambda^{2}_{a_k} \Lambda^{2}_{a_l} \left(\Gamma_{a_k a_l
a_{m}^{\prime}}^{s_i}\right)^{\ast} = \delta_{a_m a_{m}^{\prime}}.
\end{eqnarray}
The advantages of the canonical form of TPS are the same as for MPS, i.e. the canonical form provides a well-defined basis for evaluations of observables and the imaginary-time evolution.

\subsection{Imaginary-time evolution}
In order to obtain the model's ground state within this class of iTTN we evolve an initial state $\vert \psi \rangle$ in imaginary time. Since the Hamiltonian is given by the sum over nearest neighbors of products of commuting operators, we can implement the imaginary-time evolution as a product of local unitary operators using the second-order Suzuki-Trotter decomposition of the evolution operator: $e^{-H t} = \lim_{n \rightarrow \infty} \left( e^{-H \delta t} \right)^{n}$. This decomposition incurs a systematic non-accumulating error of $\mathcal{O}(\delta t ^2)$ which can be neglected if the time step $\delta t$ is sufficiently small (in our simulations we use $\delta t = 10^{-6}$ as the final time step).
\begin{figure}
\subfigure[\label{fig:UpdateStep1}]{
   \includegraphics[width=230pt]{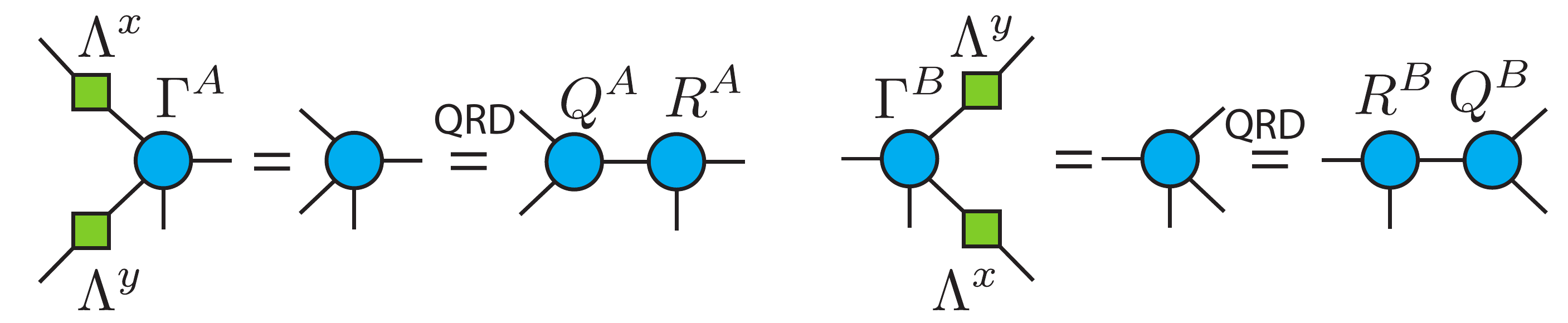}
   }
  \subfigure[\label{fig:UpdateStep2}]{
   \includegraphics[width=115pt]{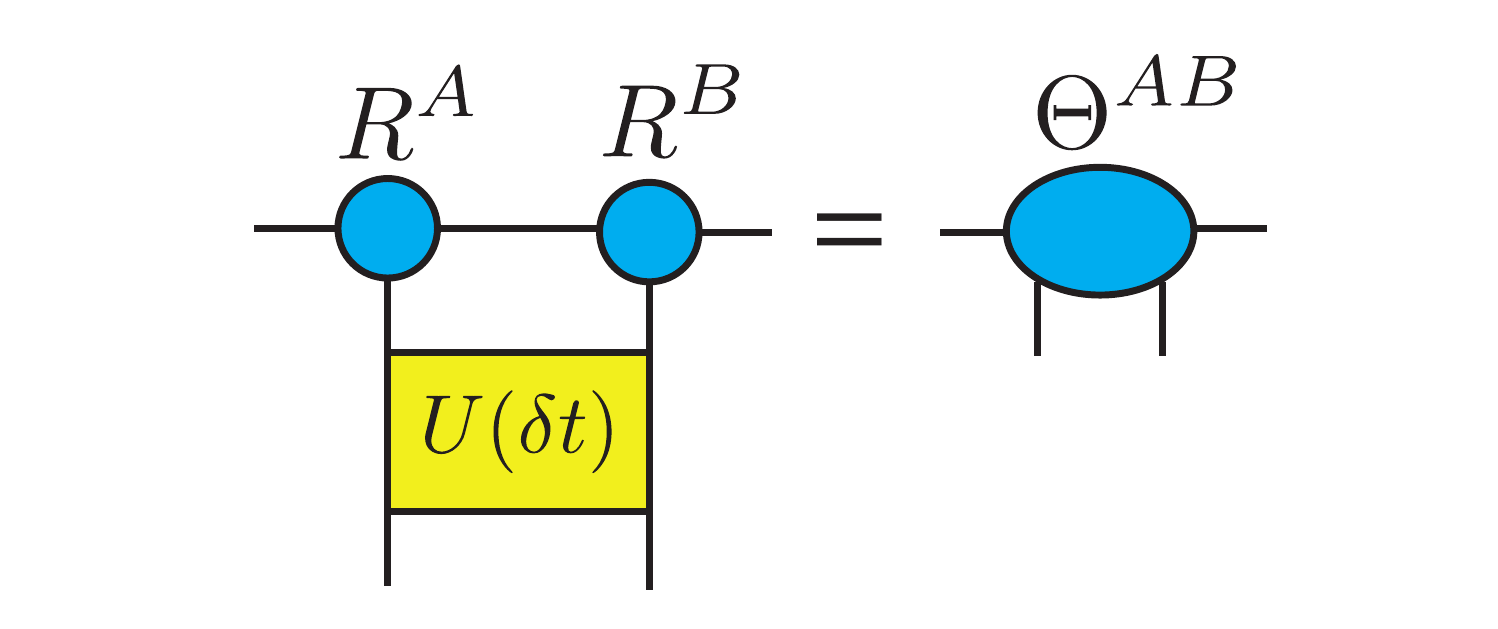}
   }
   \subfigure[\label{fig:UpdateStep3}]{
   \includegraphics[width=115pt]{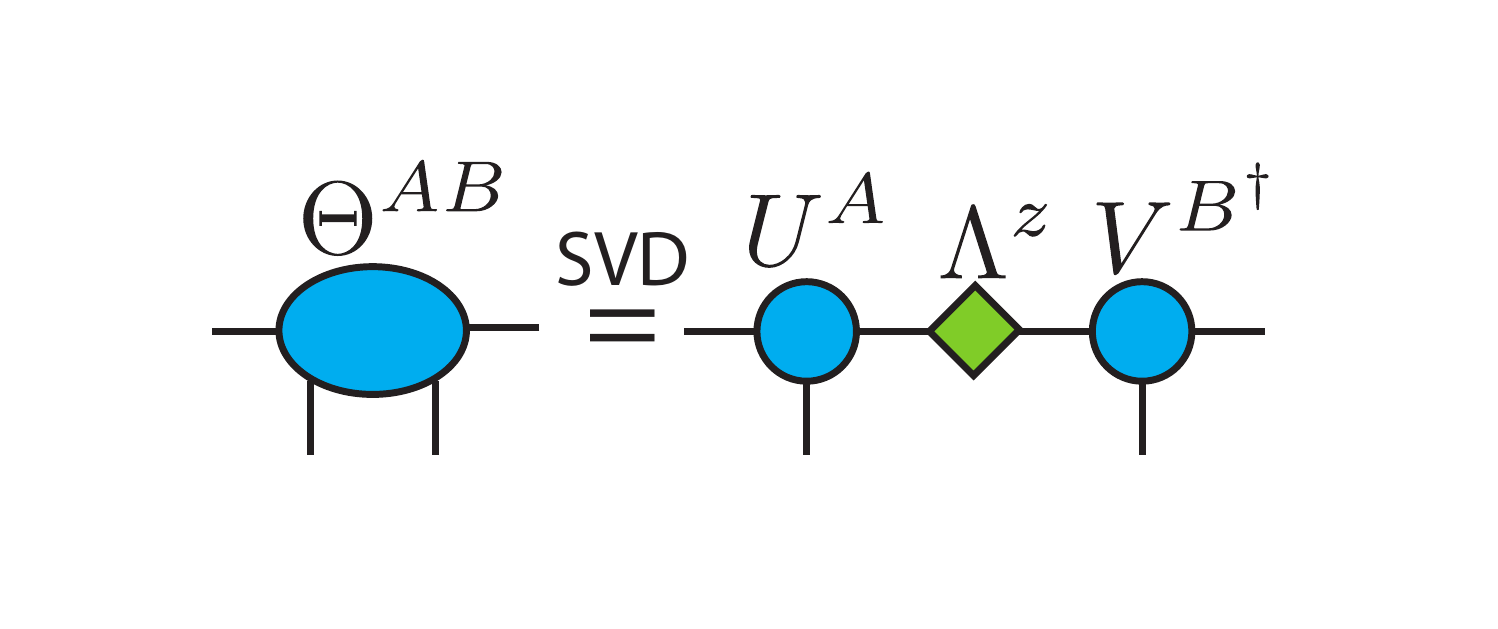}
   }
   \subfigure[\label{fig:UpdateStep4}]{
   \includegraphics[width=230pt]{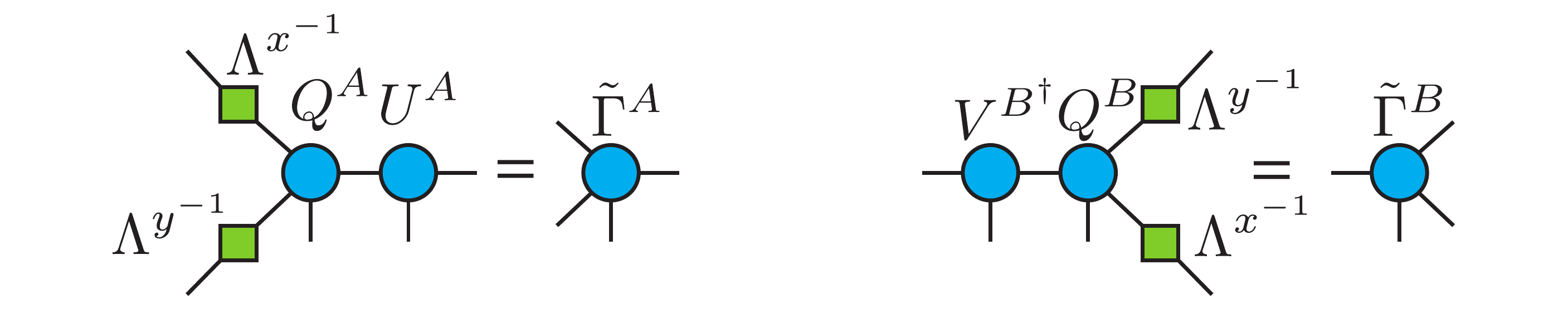}
   }
\caption{\label{fig:SimpleUpdate}The simple update procedure for a single two-site unit cell. By repeating this procedure for every tensor combination one full update step is completed. Details on the procedure are provided in the text.}
\end{figure}
While in principle each site tensor $\Gamma^{[i]}$ and bond tensor $\Lambda^{k}$ can be different for different sites and bonds we will calculate the ground state within the translation-invariant sector of iTTN.  In this picture, let us consider the effects of the imaginary-time evolution of the tree tensor network by a translationally invariant Hamiltonian. If given a translation-invariant state, the symmetry of the Hamiltonian guarantees it to never be broken in time, thus enabling us to describe the full state by examining only a small number of sites. However, if the system is in a phase in which the translational symmetry is spontaneously broken, it turns out to be advantageous to allow for a larger unit cell and to guide the wave function into a symmetry broken state.

Since the infinite Bethe lattice is self-similar, a translation-invariant ansatz state is the natural choice. For numerical reasons it is advantageous to slightly break that translational invariance by adopting a larger unit cell. Canonically a two-site unit cell is used, but here we will extend this scheme and for some calculations employ a six-site unit cell (see Fig. \ref{fig:Ansatz}), which enables us to also capture more involved states such as a dimerized state. In accordance with the canonical iTEBD algorithm, the two-site unit cell consists of two site tensors $A$ and $B$ and three bond vectors $\Lambda^{x}$, $\Lambda^{y}$, and $\Lambda^{z}$, whereas the six-site cell uses six site tensors and nine bond tensors.

We now have all the ingredients necessary to compute the ground state. By repeatedly applying the near-unitary operators
$U(\delta t) = e^{-H \delta t}$ to an initial state $\vert \psi \rangle$ and then truncating the entanglement spectrum we
can obtain the ground state. The operator's near-unitary nature allows us to perform the truncation in a well-defined basis, yielding a stable algorithm to find the ground state on the Bethe lattice. This procedure was introduced in the context of MPS as iTEBD \cite{Vidal-2007} and later on generalized the Bethe lattice.\cite{Nagaj-2008,Nagy-2012,Li-2012} The same algorithm is also used to find an approximate TPS representation on higher-dimensional lattices in the so-called \textit{simple update}.\cite{Jiang-2008} However, as the update algorithm ignores the loops present in a two-dimensional lattice, the TPS found is not optimal to represent the 2D ground state.

The update procedure for the imaginary-time evolution of a tree tensor network consisting of an infinitely repeated
two-site unit cell is now given by the following steps (Fig.~\ref{fig:SimpleUpdate}):

(1) Contract the site tensors with the adjacent bond vectors, leaving one bond open.

(2) Compute the QR decomposition of the resulting tensors relative to the open bond. This modification was recently introduced by Wang\cite{Wang-2011} to reduce the scaling of the update with the bond dimension and to stabilize the update procedure.

(3) Contract the evolution operator with the $R$ tensors resulting from the QR decomposition and calculate the singular
value decomposition of the resulting 4-index tensor.

(4) Truncate the entanglement spectrum to $\chi$ entries and absorb the unitary matrices in the $Q$ tensors.

(5) Contract the tensors with the inverse bond vectors to obtain the updated site tensors.

By repeating this procedure for every bond, one full update step is completed, i.e., the full lattice is updated.
Repeatedly applying the update to the system while decreasing the time step brings the initial wave function
increasingly closer to the true ground state wave function.

For a general Bethe lattice with coordination number $z$ the computational cost of this algorithm scales as
$\mathcal{O}(\chi^{3z-3}) + \mathcal{O}(d^2 \chi^{z+1})$ and is dominated by the cost of the SVD step. The high cost of
the SVD limits our algorithm to values of $\chi \approx 40$ on a desktop machine when no additional symmetries are used.

\subsection{Symmetries}
The algorithm's performance can be improved by exploiting the model's symmetries, which enable a decomposition of the matrices into block-diagonal matrices, hence reducing the cost of the numerical operations. Here, we made use of the models $U(1)$ symmetry in the $S_z$ sector. By implementing this symmetry the computational cost can be significantly reduced, allowing a larger cut-off dimension of $\chi \approx 80$. Details of how to implement this symmetry can be found in e.g. Refs. \onlinecite{Singh-2009,Singh-2012a}.

\section{Symmetry protected topological phases} \label{sec:sptp}
Symmetry protected topological phases (SPTP) are gapped phases which cannot be characterized by any local order parameters and are distinct from trivial phases (i.e., product states) only in the presence of certain symmetries. In a series of works it had been shown that these phases can be completely  characterized using  projective representations of the symmetries present.\cite{Gu-2009,Pollmann-2010,Fidkowski-2010,Turner-2010,Pollmann-2012,ChenGu-2011,ChenGu-2011-2,Schuch-2011} In spin systems this means physically  that the spin fractionalizes and the projective representation is due to localized spin-half degrees at the edge of a cut. We briefly review SPTP  for one-dimensional systems  and  show that the concept directly generalizes to the Bethe lattice. 

\begin{figure}
\includegraphics[width=230pt]{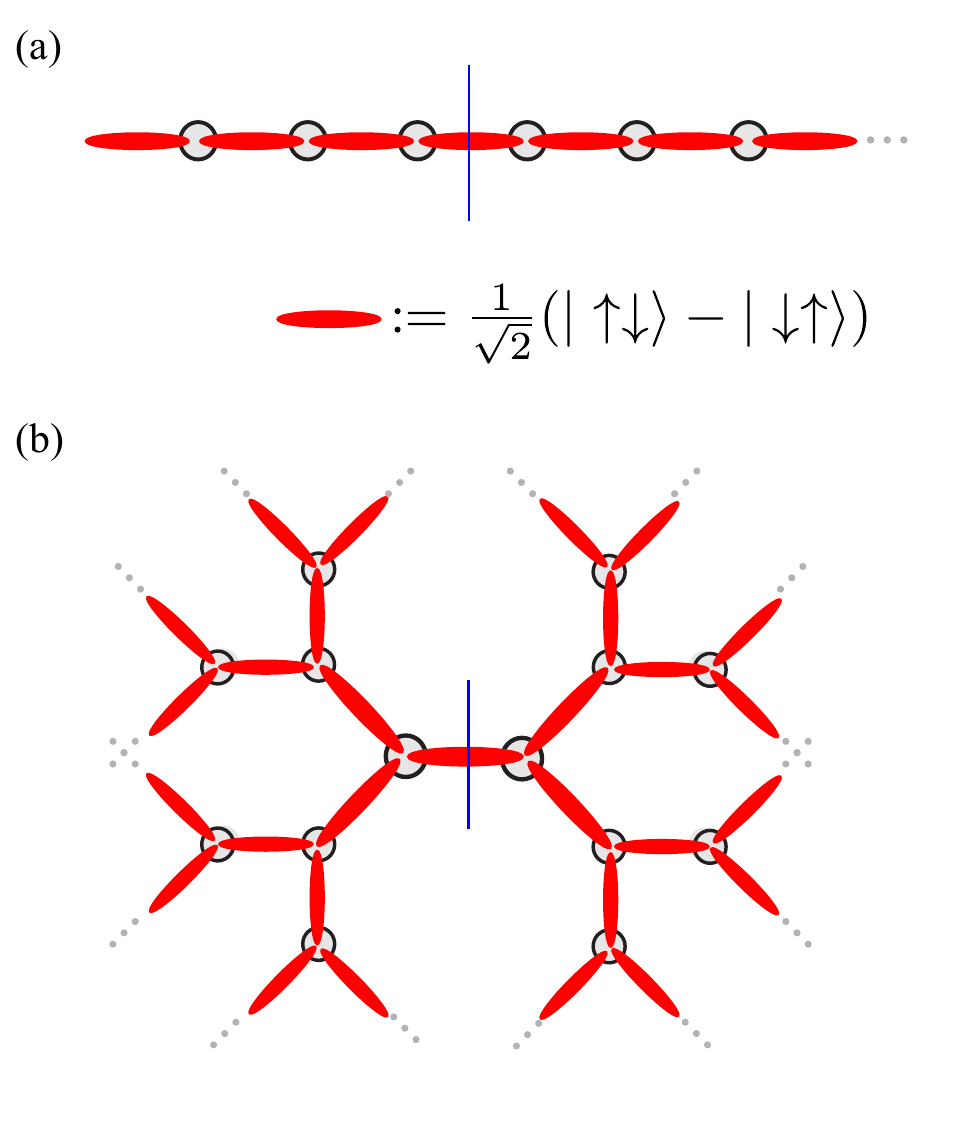}
\caption{\label{fig:SPTP} Diagrammatic representation of the AKLT wave function on a chain (a) and a $z=3$ Bethe lattice (b). The red ovals are representing $S=1/2$ singlets. A Schmidt decomposition at the blue line cuts in both cases one singlet, leaving behind localized $S=1/2$ edge spins.}
\end{figure}
We start from a state $|\psi \rangle $ on an infinite chain that is invariant under an \emph{internal symmetry}. The internal symmetry is represented in the spin basis by a unitary matrix $\Sigma$ acting on each site so that   $|\psi\rangle \rightarrow \left[ \bigotimes_i \Sigma(i)\right]|\psi\rangle$. We  perform a Schmidt decomposition of the system into two subsystems (see Fig.~\ref{fig:SPTP}a) by cutting one bond. We now only consider   the \emph{important} Schmidt states  which correspond to Schmidt values $\Lambda_{\alpha}~>~\epsilon$ for a given  $\epsilon~>~0$. These Schmidt states transform under a symmetry transformation as (modulo an overall phase):
\begin{equation}
\left[ \bigotimes_i \Sigma(i)\right]|\alpha R\rangle = \sum_{\alpha^{\prime}} U_{\alpha\alpha^{\prime}}|\alpha^{\prime} R\rangle,
\end{equation}
where $U$ is a unitary matrix which commutes with the $\Lambda $ matrices.\cite{Pollmann-2010,PerezGarcia-2008} Similarly, the left Schmidt states $|\alpha L\rangle$ transform by the conjugate matrix. As
the symmetry element $g$ is varied over the whole group, a set of matrices $U_g$ results. The matrices $U_g$ form a $\chi-$dimensional (projective) representation
of the symmetry group. A projective representation is like an
ordinary regular representation up to phase factors; i.e., if  $\Sigma^g\Sigma^h=\Sigma^{gh}$, 
then 
\begin{equation}
U_{g}U_{h}=e^{i\rho(g,h)}U_{gh}.
\label{eq:rho}
\end{equation}
The phases $\rho(g,h)$ can be used to classify different 
topological phases.\cite{Pollmann-2010,Pollmann-2012,Schuch-2011,Chen-2011}  

As a concrete example, we now consider the Haldane phase around the AKLT state in the presence of a $\mathds{Z}_2\times \mathds{Z}_2$ symmetry. The generators of the symmetry group are the spin rotations $\mathcal{R}_x=\exp(i\pi S^x)$ and $\mathcal{R}_z=\exp(i\pi S^z)$. The phases for each spin rotation individually (e.g., $U_{x}^2=e^{i\alpha}\mathds{1}$) can 
be removed by redefining the phase of the corresponding $U$-matrix.  However, the representations of 
$\mathcal{R}_x\mathcal{R}_z$ and $\mathcal{R}_z\mathcal{R}_x$ can also differ by a phase, which it turns out must be $\pm 1$:
\begin{equation}
U_{x}U_{z}=\pm U_{z}U_{x}.
\label{eq:schur}
\end{equation}
I.e., the matrices either commute or anti-commute. This resulting  phase cannot be gauged away because the phases 
of $U_{x}$ and $U_{z}$ enter both sides of the equation in the same way. Thus we have two different classes of projective representations.  If the phase is $-1$, then the spectrum
of $\Lambda $ is doubly degenerate, since $\Lambda $ commutes with the two
unitary matrices $U_{x}$, $U_{z}$ which anti-commute among themselves. For
the AKLT state considered here, the Schmidt states have  half-integer edge spins (see  Fig.~\ref{fig:SPTP}). Thus we find  $U_{x}=\sigma _{x}$ and $U_{z}=\sigma _{z}$, therefore $%
U_{x}U_{z}=-U_{z}U_{x}$, and the Haldane phase is protected if the system is
symmetric under both $\mathcal{R}_{x}$ and $\mathcal{R}_{z}$. An analogous argument can be made for inversion symmetry (i.e., spatial inversion of the system at a bond) and time reversal symmetry.\cite{Pollmann-2010,Chen-2011}  

Using the above arguments, we can now characterize the $S=3/2$ AKLT state on the $z=3$ lattice. 
As illustrated in Fig.~\ref{fig:SPTP}b, a single cut through a bond separates the Bethe lattice into two disconnected subsystems. 
In the AKLT state, any of the bonds has a $S=1/2$ singlet and thus the Schmidt states have localized half-integer spins at the edges. 
Thus the transformation of the Schmidt states under a symmetry operation  yields a (projective) representation $U$ of the symmetry group which characterizes the phase. 
Note that there are in fact different kinds of SPTP that can be realized on the Bethe lattice depending on the fractionalization of the spin. 
For example, in an $S=1$ model on a $z=3$ lattice with strong dimerization (two strong bonds and one weak bond), we would obtain a network of $S=1$ Haldane chains. 

\section{Numerical Results} \label{sec:results}
We employ our variant of the iTEBD algorithm to study the bilinear-biquadratic-bicubic Heisenberg model defined in Eq.~(\ref{eq:model}) over the full parameter range of $\varphi = -\pi \ldots \pi$, $\theta = -\frac{\pi}{2} \ldots \frac{\pi}{2}$. To ensure unbiased results we use different iTEBD implementations, one with a two-site unit cell and one with a six-site unit cell (see Fig.~\ref{fig:Ansatz}). The simulations were conducted both with and without explicitly conserving the model's $U(1)$ symmetry and were started from different initial wavefunctions ranging from completely random to fully polarized initial states. Furthermore, we also studied the dependence on the evolution scheme by starting the imaginary-time evolution with a slightly modified evolution operator (e.g. by adding a small ferromagnetic or symmetry-breaking interaction term to the Hamiltonian) and only later in the calculation using the actual Hamiltonian to converge the trial wavefunction to the groundstate. 

The simulations were started with a large time step of $\delta t = 0.1$ which was then in several steps decreased to $\delta t = 10^{-6}$, reducing the Trotter error to insignificance. These checks are commonly accepted best practice for any imaginary-time evolution, as the overlap of the initial state and the true groundstate has to be finite in order for the algorithm to be able to evolve the trial wavefunction to the groundstate. Hence we have to ascertain the result's independence of the procedure and the initial states. As expected we find the strongest dependence on the initial tensors in the ferromagnetic (anti-ferromagnetic) phase when starting with an anti-ferromagnetic (ferromagnetic) initial state, but no significant correlation otherwise. To establish that our findings are not dependent on the unit cell, i.e., the constraint of a two-site unit cell, we also implemented a variant of the iTEBD algorithm that operates on a six-site unit cell, see Fig.~\ref{fig:2siteAnsatz}. This implementation enables us to also describe a dimerized phase which would otherwise be able to escape characterization as it is not commensurable with a two-site unit cell.

\begin{figure}
  \includegraphics[height=180pt]{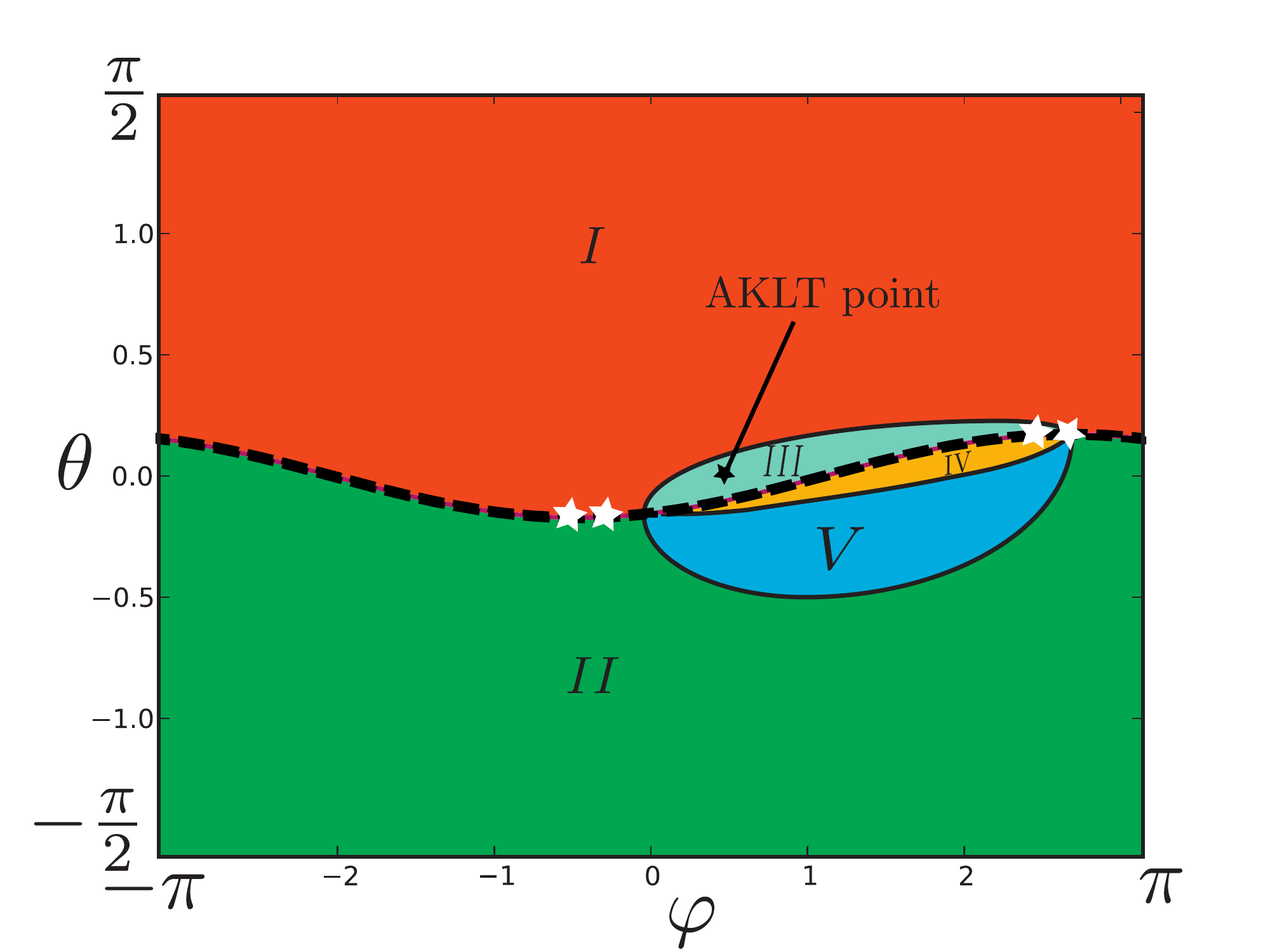}
  \caption{\label{fig:phases}The phase diagram as determined by the tree network iTEBD method. We established the presence of an anti-ferromagnetic phase ($I$), a ferromagnetic phase ($II$), the Haldane phase ($III$), a dimerized phase ($IV$), and a ferrimagnetic phase ($V$). For reference purposes the phase diagram also includes the AKLT point denoted by a black star and the $SO(5)$ symmetric line denoted by the dotted line between phases $I$ and $II$. The $SU(4)$ symmetric points are marked by white stars and lie on the $SO(5)$ line.}
\end{figure}

Special attention was paid to the four $SU(4)$ symmetric points shown in the phase diagram Fig. ~\ref{fig:phases} where we ran a variety of simulations to ascertain the model's behavior. With the exception of the multi-critical point at $\varphi = 2.93$, $\theta = 0.17$ we were not able to observe any behavior diverging from the encompassing phase. 

Our results for the phase diagram of the bilinear-biquadratic-bicubic Heisenberg model are shown in Fig.~\ref{fig:phases}. We find five different phases: a N\'{e}el phase, a ferromagnetic polarized phase, the Haldane phase, a dimerized phase, and a ferrimagnetic phase.

\subsection{(Anti-)Ferromagnetic phases}
\begin{figure}
  \includegraphics[height=180pt]{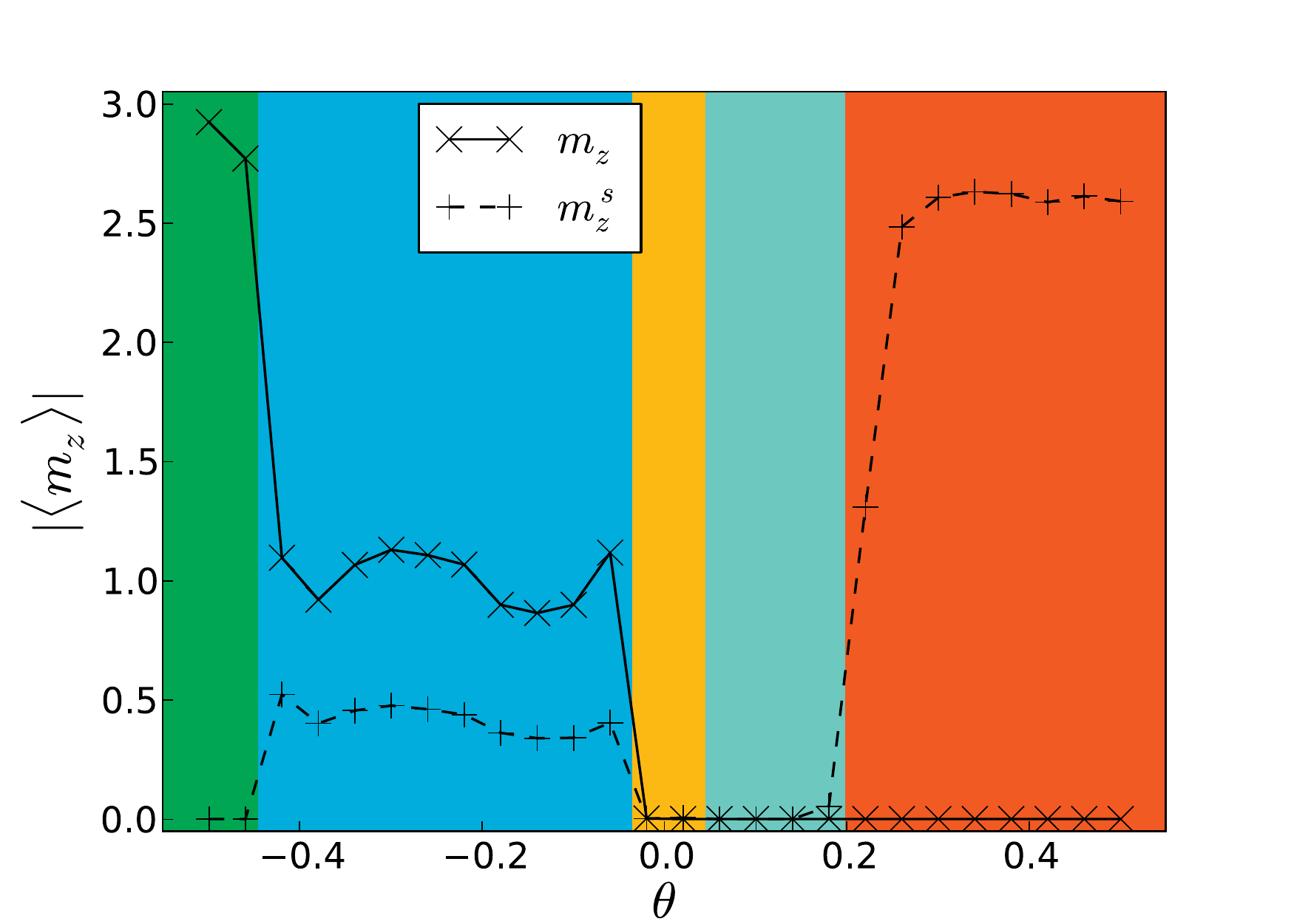}
  \caption{\label{fig:cut}The (staggered) magnetization for various of values of $\theta$ and $\varphi = 0.9$ with $\chi=24$. The ferromagnetic phase can be identified as the region with maximal uniform magnetization per site, whereas the anti-ferromagnetic phase is distinguished by the finite staggered magnetization and vanishing uniform magnetization per site. Inbetween those two phases are the dimerized and the Haldane phases, where both the staggered  ($m_{z}^{s}$) and uniform  magnetization ($m_{z} $) vanish. In the ferrimagnetic phase they take intermediate values. The colors correspond to the phases introduced in Fig. \ref{fig:phases}.}
\end{figure}

We start our analysis of the phase diagram with the straightforward phases, i.e. the AFM and FM phases. First we consider the staggered magnetization $m_{z}^{s}$
and the uniform magnetization $m_{z}$.
The polarized phases can easily be identified by observing the two different order parameter's behavior shown in Fig.~\ref{fig:cut}. In the ferromagnetic region $\theta < -\frac{\pi}{3}$ the staggered magnetization $m^{s}_{z}$ disappears, while the magnetization per site $m_z$ is maximal in this region. The opposite holds for the anti-ferromagnetic part of the phase diagram, where only the staggered magnetization remains non-zero.
Inspired by the (anti-) nematic order found in mean-field studies\cite{Fridman-2011} we also calculate the octupolar order parameter in the vicinity of the $SO(5)$ line. Again we observe the absence of different order, except for some small contributions in the ferromagnetic region at negative $\varphi$. This might be due to remnants of the classical order persisting to zero temperature.

\subsection{Dimerized phase}

\begin{figure}
  \includegraphics[height=180pt]{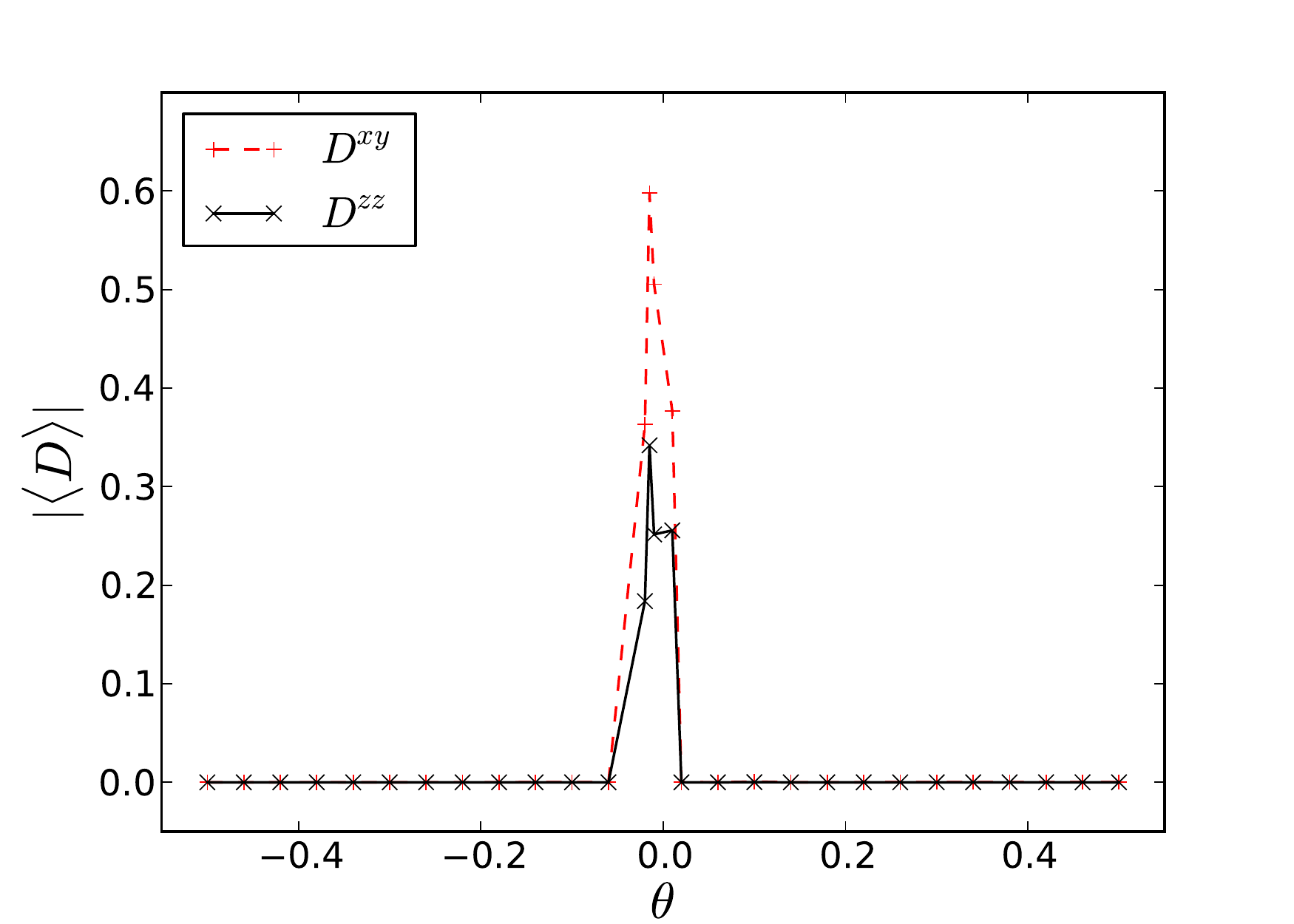}
  \caption{\label{fig:dimers}Behavior of the dimer order parameters defined in Eq.~\ref{eq:dimerorder} for a cut through the phase diagram at $\varphi = 0.9$ with $\chi = 24$. The increase of the dimer order parameters occurs at the boundary between the Haldane and the ferrimagnetic phase.}
\end{figure}

We determine  a dimerized phase to be present in the region denoted by $IV$ in the phase diagram (Fig.~\ref{fig:phases}). This phase is hard to characterize due to its diminutive size and vanishing magnetization, however,  careful calculations strongly indicate its existence. To  determine the properties of this elusive plains we calculate the $xy$ and $z$ components of the dimer order parameters $D^{xy}$ and $D^{xy}$ defined via
\begin{eqnarray} \label{eq:dimerorder}
  D^{xy}_{i, j, k} =& \langle \left( S^{x}_{i}S^{y}_{j}+S^{y}_{i}S^{x}_{j}\right) \\ &- \left(S^{x}_{j}S^{y}_{k}+S^{y}_{j}S^{x}_{k}\right) \rangle \\
  D^{z}_{i, j, k} =& \langle S^{z}_{i}S^{z}_{j}-S^{z}_{j}S^{z}_{k}\rangle,
\end{eqnarray}
where $i$, $j$, and $k$ label consecutive lattice sites residing on different shells (e.g. sites $A$, $B$, and $C$ in Fig.~\ref{fig:Ansatz} (b)).
Calculation of these order parameter components reveals them to vanish for the magnetically ordered phases, as well as in the Haldane phase. Only in the dimer phase do they assume finite values. The hypothesis of a dimer phase is further corroborated by the vanishing magnetization.
As opposed to the Haldane phase we also fail to observe finite edge spins in this phase.
Together these observations indicate the existence of  a narrow dimerized phase close to the $SO(5)$ symmetric line. 

\subsection{Ferrimagnetic phase}
We found a ferrimagnetic phase that exists between the dimer phase and the ferromagnetic phase. This phase shows both a finite staggered magnetization as well as a finite magnetization per site, but displays vanishing dimer order parameters. 
\begin{figure}
  \includegraphics[height=180pt]{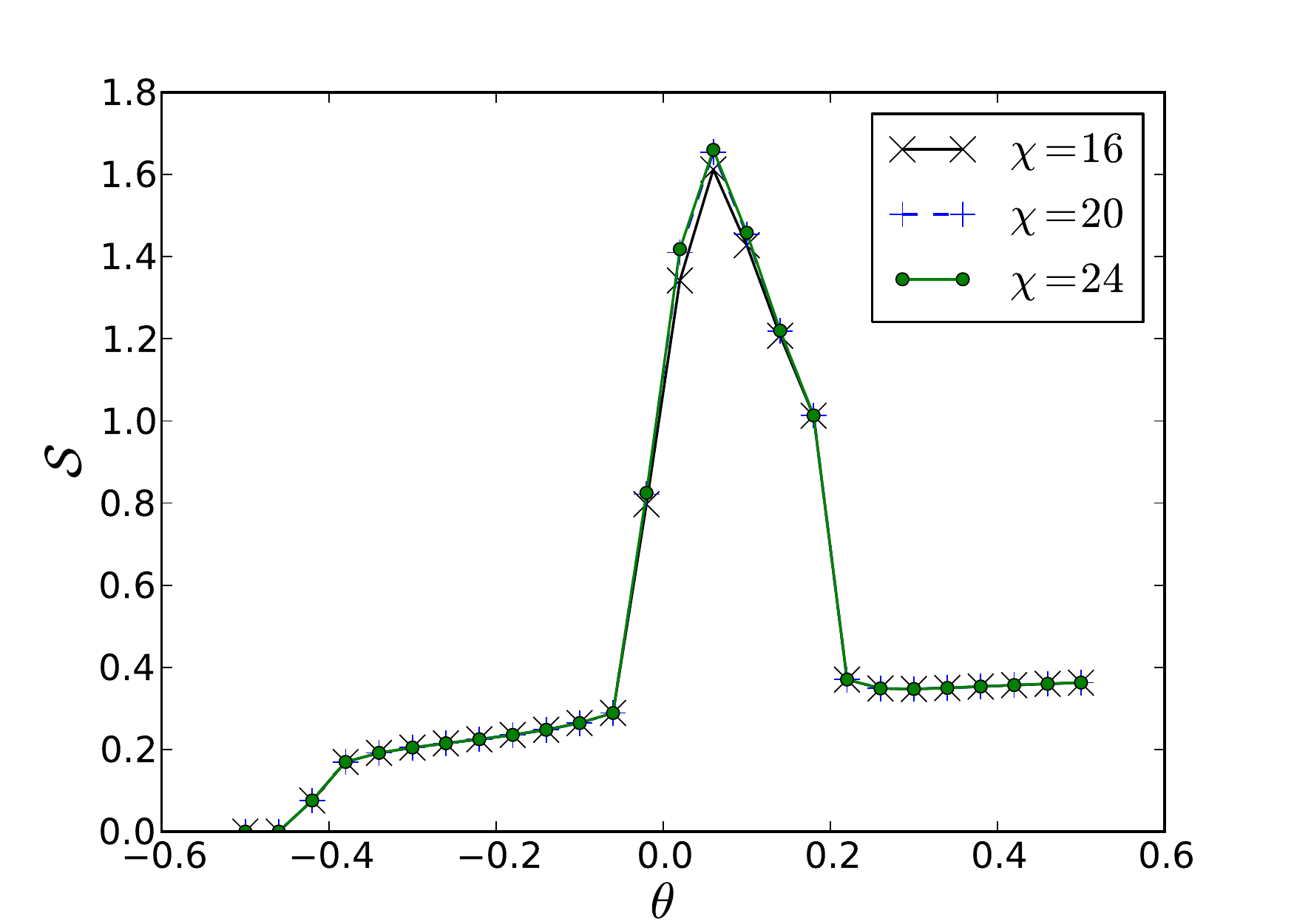}
  \caption{\label{fig:Entropy}The entropy for a cut in $\theta$ direction through the phase diagram at $\varphi = 0.9$ for various values of $\chi$ showing the entropy to be clearly converged when increasing the bond dimension.}
\end{figure}
As a test for this phase we try adding a small (staggered) field in the $z$ direction to the Hamiltonian. Performing the simulation with this modified Hamiltonian results in strong polarization which could be both anti-ferromagnetic or ferromagnetic depending on the applied field. 

\subsection{Haldane phase}
\begin{figure}
  \includegraphics[height=180pt]{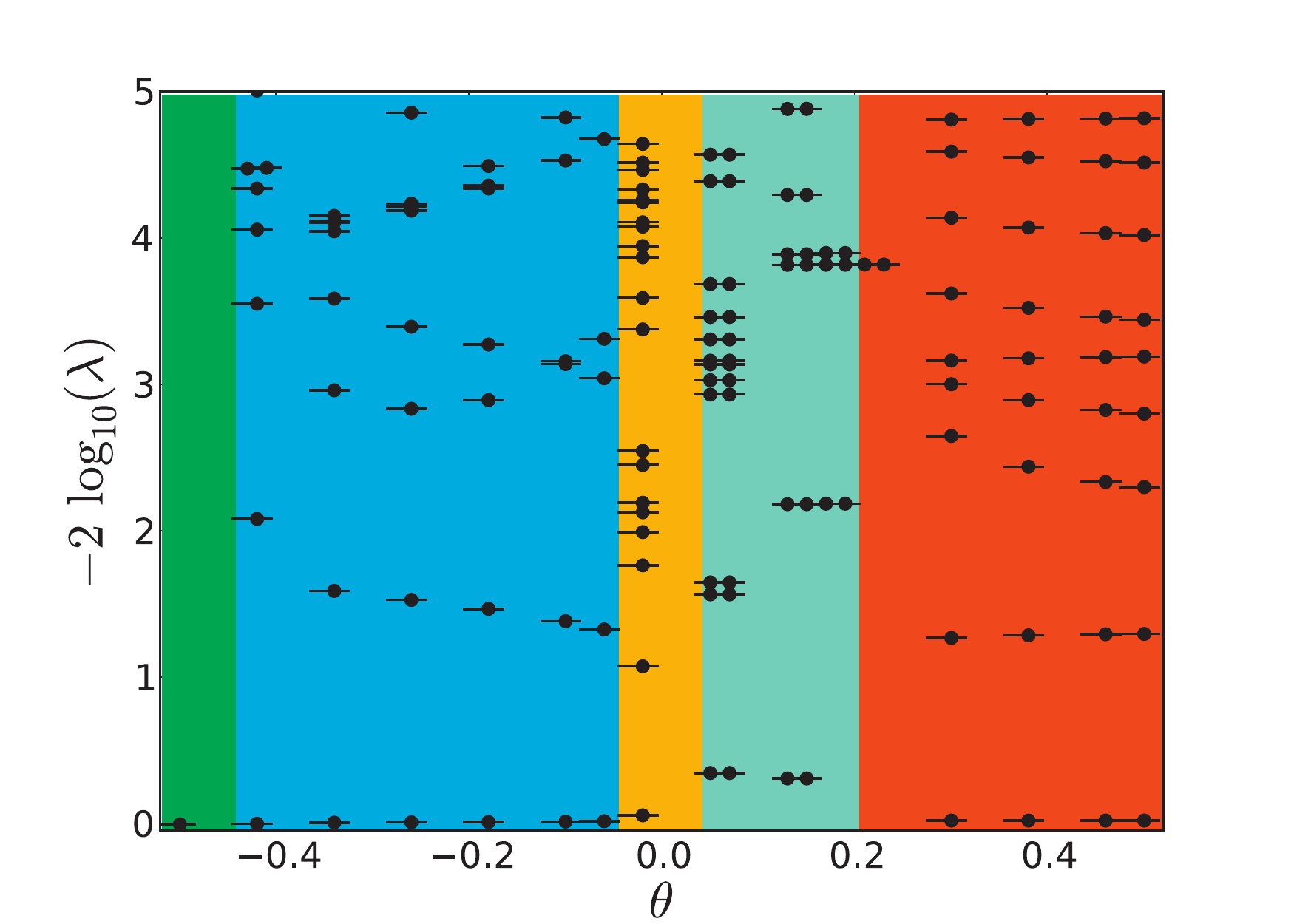}
  \caption{\label{fig:entanglement}Entanglement spectrum throughout a cut in the phase diagram at $\varphi = 0.9$ with $\chi = 24$ for various values of $\theta$. Clearly visible is the doubling of the levels in the Haldane phase.}
\end{figure}
The last phase is a phase with exponentially decaying correlations and no broken symmetries  (i.e., no local order parameter exists) which is identified as a SPTP. In this phase, the entanglement spectrum always displays an even degeneracy in the entire phase, which is clearly visible in the iTEBD calculations. The identification of this phase  rests on the presence of a finite $S=1/2$ edge spin and its characteristic degeneracies in the entanglement spectrum shown in Fig.~\ref{fig:entanglement}.  As a direct evidence, we also calculated the edge spin of the Schmidt states directly and find a localized spin-1/2. In our simulations we are also able to observe the AKLT point where the ansatz state reduces to an exact tensor network with bond dimension $\chi = 2$. By adding small perturbations which destroy all the necessary symmetries to protect the phase, it is possible to drive the system out of the Haldane phase  without a phase transition (i.e., the degeneracies in the spectrum are lifted for an arbitrarily small perturbation). We also checked numerically  that   the phase is robust against small perturbations which do not break the symmetries needed to protect it. All of these observation can be  explained by the presence of a SPT phase as discussed in section ~\ref{sec:sptp}.

\section{Summary\label{sec:sum}}

In this work, we have studied a general $SU(2)$ symmetric spin-3/2 model on the z = 3 Bethe lattice using the infinite Time Evolving Block Decimation (iTEBD) method. We found that the model exhibits a rich phase diagram containing several magnetic phases, a dimerized phase as well as a symmetry protected topological phase (SPTP). The magnetic phases were identified by calculating the uniform and the staggered magnetization. We found a polarized ferromagnetic phase, an anti-ferromagnetic as well as a ferrimagnetic phase with finite (staggered) magnetization. Our simulations suggest the presence of a dimerized phase with vanishing magnetization and finite dimer order parameters. We also identified a symmetry protected topological phase which shows all the key features of the Haldane phase. This phase is characterized by spin-1/2 edge spins and degeneracies in the entanglement spectrum. 

\section*{Acknowledgment}
We thank Erez Berg, Alexei Kolezhuk, Karlo Penc, Stephan Rachel, and Ari Turner for stimulating discussions. SD would like to thank the Max Planck Institute for the Physics of Complex Systems for the generous access to their computing facilities.
\bibliographystyle{apsrev1}

%

\end{document}